# Direct-coupling analysis of residue co-evolution captures native contacts across many protein families

Faruck Morcos [1,a], Andrea Pagnani [1,b,c], Bryan Lunt [a], Arianna Bertolino [d], Debora S. Marks [e], Chris Sander [f], Riccardo Zecchina [b,c], José N. Onuchic [a,g], Terence Hwa [a], Martin Weigt [b,h]

[a] Center for Theoretical Biological Physics, University of California at San Diego, La Jolla, CA 92093-0374; [b] Human Genetics Foundation, Via Nizza 52, 10126 Torino, Italy; [c] Center for Computational Studies and Dipartimento di Fisica, Politecnico di Torino, Corso Duca degli Abruzzi 24, 10129 Torino, Italy; [d] Institute for Scientific Interchange, Viale S. Severo 65, 10133 Torino Italy; [e] Department of Systems Biology, Harvard Medical School, 200 Longwood Ave., Boston, MA 02115; [f] Memorial Sloan-Kettering Cancer Center, Computational Biology Center, 1275 York Avenue, New York, NY 10065; [g] Center for Theoretical Biological Physics, Rice University, Houston, TX 77005-1827; [h] Laboratoire de Génomique des Microorganismes, UMR 7238, Université Pierre et Marie Curie, 15 rue de l'École de Médecine, 75006 Paris, France

**Corresponding authors email:** José N. Onuchic, jonuchic@ucsd.edu; Terence Hwa, hwa@ucsd.edu; Martin Weigt, martin.weigt@upmc.fr

**Classification:** BIOLOGICAL SCIENCES: Biophysics and Computational Biology

**Abstract:** The similarity in the three-dimensional structures of homologous proteins imposes strong constraints on their sequence variability. It has long been suggested that the resulting correlations among amino acid compositions at different sequence positions can be exploited to infer spatial contacts within the tertiary protein structure. Crucial to this inference is the ability to disentangle direct and indirect correlations, as accomplished by the recently introduced Direct Coupling Analysis (DCA) (Weigt *et al.* (2009) *Proc Natl Acad Sci* 106:67). Here we develop a computationally efficient implementation of DCA, which allows us to evaluate the accuracy of contact prediction by DCA for a large number of protein domains, based purely on sequence information. DCA is shown to yield a large number of correctly predicted contacts, recapitulating the global structure of the contact map for the majority of the protein domains examined. Furthermore, our analysis captures clear signals beyond intra-domain residue contacts, arising, e.g., from alternative protein conformations, ligand-mediated residue couplings, and inter-domain interactions in protein oligomers. Our findings suggest that contacts predicted by DCA can be used as a reliable guide to facilitate computational predictions of alternative protein conformations, protein complex formation, and even the *de novo* prediction of protein domain structures, provided the existence of a large number of homologous sequences which are being rapidly made available due to advances in genome sequencing.

---

[1] These authors contributed equally

\body

## Introduction

Correlated substitution patterns between residues of a protein family have been exploited to reveal information on the structures of proteins (1-10). However, such studies require a large number (e.g., the order of one thousand) of homologous yet variable protein sequences. In the past, most studies of this type have therefore been limited to a few exemplary proteins for which a large number of such sequences happened to be already available. However, rapid advances in genome sequencing will soon be able to generate this many sequences for the majority of common bacterial proteins (11). Sequencing a large number of simple eukaryotes such as yeast can in principle generate similar number of common eukaryotic protein sequences. In this paper, we provide a systematic evaluation of the information contained in correlated substitution patterns for predicting residue contacts, a first step towards a purely sequence-based approach to protein structure prediction.

The basic hypothesis connecting correlated substitution patterns and residue-residue contacts is very simple: If two residues of a protein or a pair of interacting proteins form a contact, a destabilizing amino-acid substitution at one position is expected to be compensated by a substitution of the other position over the evolutionary time scale, in order for the residue pair to maintain attractive interaction. To test this hypothesis, the bacterial two-component signaling (TCS) proteins (12) have been used due to the large number of TCS protein sequences, which numbered in the thousands already 5 years ago (13). Simple co-variance based analysis was first applied to characterize interactions between residues belonging to partner proteins of the TCS pathways (14, 15); it was found to partially predict correct inter-protein residue contacts, but also many residue pairs which are far apart. A major shortcoming of co-variance analysis is that correlations between substitution patterns of interacting residues induce secondary correlations between non-interacting residues. This problem was subsequently overcome by the *Direct Coupling Analysis* (DCA) (16, 17), which aims at disentangling direct from indirect correlations. The top 10 residue pairs identified by DCA were all shown to be true contacts between the TCS proteins, they were used to guide the accurate prediction

(3Å RMSD) of the interacting TCS protein complex (18, 19). Furthermore, DCA was used to shed light on interaction specificity and inter-pathway crosstalk in bacterial signal transduction (20).

Due to rapid advances in sequencing technology, there exists by now a large number of bacterial genome projects, approximately 1700 completed and 8300 ongoing (11). These genome sequences can be used to compute correlated substitution patterns for a large number of common bacterial proteins and interacting protein pairs, even if they are not duplicated, i.e., present at one copy per genome on average. DCA can then be used in principle to infer the interacting residues and eventually predict tertiary and quaternary protein structures for the majority of bacterial proteins, as has been done so far for the TCS proteins. Here we address a critical question for this line of pursuit – how well does DCA identify native residue contacts in proteins other than TCS?

Previously, a message-passing algorithm was used to implement DCA (16). This approach, here referred to as mpDCA, was rather costly computationally, since it is based on a slowly converging iterative scheme. This cost makes it unfeasible to apply mpDCA to large scale analysis across many protein families. Here we will introduce a new algorithm, mfDCA, which is based on the mean-field approximation of DCA. mfDCA is $10^3$-$10^4$-times faster than mpDCA, and hence can be used to analyze many long protein sequences rapidly. By analyzing 131 large domain families for which accurate structural information is available, we show that mfDCA captures a large number of intra-domain contacts across these domain families. Together, the predicted contacts are able to recapitulate the global structure of the contact map. Many cases where mfDCA finds strong correlation between distant residue pairs, have interesting biological reasons, including inter-domain contacts, alternative structures of the same domain, and common interactions of residues with a ligand. The mfDCA results are found to outperform those generated by simple co-variance analysis as well as a recent approximate Bayesian analysis (10).

## Results and Discussion

### A fast DCA algorithm

In this study, we wish to characterize the correlation between the amino-acid occupancy of residue positions as a predictor of spatial proximity of these residues in folded proteins. Starting with a multiple sequence alignment (MSA) of a large number of sequences of a given protein domain, extracted using Pfam's Hidden Markov Models (HMMs) (21, 22), the basic quantities in this context are the frequency count $f_i(A)$ for a single MSA column $i$, characterizing the relative frequency of finding amino acid $A$ in this column, and the frequency count $f_{ij}(A,B)$ for pairs of MSA columns $i$ and $j$, characterizing the frequency that amino acids $A$ and $B$ co-appear in the same protein sequence in MSA columns $i$ and $j$. Alignment gaps are considered as the 21$^{st}$ amino acid. Mathematical definitions of these counts are provided in Methods.

The raw statistical correlation obtained above suffers from a *sampling bias,* resulting from phylogeny, multiple-strain sequencing, and a biased selection of sequenced species. The problem has been discussed extensively in the literature (10, 23-26). In this study, we implemented a simple sampling correction, by counting sequences with more than 80% identity and re-weighting them in the frequency counts. All the frequency calculations and results reported below are obtained using this sampling correction, the number of non-redundant sequences is measured as the effective sequence number $M_{eff}$ after reweighting, cf. Methods. The comparison to results without reweighting and to reweighting at 70% in Fig. S1 shows that reweighting systematically improves the performance of DCA, but results are robust with respect to precise value of reweighting.

A simple measure of correlation between these two columns is the mutual information (MI), defined by Eq. (3) in Methods. As we will show, the MIs turn out to be unreliable predictors of spatial proximity. Central to our approach is the disentanglement of direct and indirect correlations. This is attempted via DCA, which takes the full set of $f_i(A)$ and $f_{ij}(A,B)$ as inputs, and infers "direct statistical couplings", which generate the empirically measured correlations. Their strength is quantified by the *direct information* (DI) for each pair of MSA columns; see Eq. (12) in Methods and Ref. (16). However, the message-passing algorithm used to implement DCA in Ref. (16), mpDCA, was computationally intensive, thus limiting its use in large-scale studies. Here we developed a much faster heuristic algorithm based on a *mean-field approach*; see Methods. The new algorithm, termed mfDCA, is able to perform DCA for alignments of up to about 500 amino acids per row, as compared to 60-70 amino acids in the message passing approach. For the same protein length, mfDCA is about $10^3$-$10^4$-times faster, which results mainly from the fact that the costly iterative parameter learning in mpDCA can be solved analytically in a single step in mfDCA. This enabled us to systematically analyze hundreds of protein domains, and examine the extent to which a high DI value is a predictor of spatial proximity in a folded protein. Many residue-position pairs, which are close neighbors along the sequence, show also high MI and/or DI. To evaluate *non-trivial* predictions we therefore restricted our analysis throughout the paper to pairs, which are separated by at least 5 positions along the protein's backbone.

**Intra-domain contacts**

We shall first illustrate the correlation between the DI values and the spatial proximity of residue pairs through a specific example, namely the domain family homologous to the DNA-recognition domain (Region 2) of the bacterial Sigma-70 factor (Pfam ID PF04542). mfDCA was used to compute the DI values using $M_{eff}$~3700 non-redundant sequences, i.e., below a threshold of 80% sequence identity. The MSA columns with the 20 largest DI and MI values are mapped to the sequence of the SigmaE factor of *E. coli* (encoded by *rpoE*) whose structure has been solved to 2 Å resolution (PDB ID 1or7, (27)). The residue pairs with the 20 highest ranked DI values are connected by bonds of different colors in Fig. 1A. Those residue pairs with minimum atomic distances < 8Å are defined as "contacts" and are shown in red, the others in green[*]. As only one out of the top 20 DI pairs is green, DI is seen as a good predicator of spatial contact, characterized by a true positive (TP) rate of 95% for this protein. A similar analysis using the 20 highest MI values (Fig. 1B) yielded 13 contacts (TP=65%), illustrating a reduced predictive power by the simple co-variance analysis. Furthermore, we see that the DI predictions are more evenly distributed over the entire domain, whereas many of the MI predictions are associated with a few residues; this difference is significant for contact map prediction and will be elaborated upon below.

In order to test the generality of the predictive power of DI ranking as contacts, we applied the above analysis to 131 predominantly bacterial domain families (with > 90% of the sequences belonging to bacterial organisms). These families were selected according to the following two criteria, cf. Methods for details: (i) The family contains $M_{eff} > 1000$ non-redundant sequences after applying sampling correction for > 80% identity, in order to ensure statistical enrichment, and (ii) there exist at least two available high-quality X-ray crystal structures (independent PDB entries of resolution < 3Å), so that the degree of spatial proximity between each residue pair can be evaluated. The selected domain families encompassed a total of 856 different PDB structures (see Table S1). Note that $M_{eff}$ is found to be typically in the range of 1/3 to 1/2 of the total sequence number $M$, cf. Fig. S2.

---

[*] The choice of the relatively large value of 8Å minimum atom distance as a cutoff value for contacts is motivated later in the discussion of Fig. 2B, where the distance distribution of the top DI-pairings is analyzed.

We computed the DI values for each residue pair of the 131 domain families, and evaluated the degree to which high-ranking DI pairs corresponded to actual contacts (minimum atomic distances < 8Å), based on the available structures for each domain. The results are shown in Fig. 2A (black ✶ symbol). The x-axis represents the number of top-ranked DI pairs (separation > 5 positions along the sequence) considered and the y-axis is the average fraction of pairs up to this DI-ranking that are true contacts. The latter was calculated using the best-predicted structure[†], i.e. the PDB structure with the highest TP value, for each of the 131 families. Similar results were obtained when considering all the available structures; see below. In contrast, results computed using MI-ranking (red O symbol) gave significantly reduced TP rates[‡]. Also shown in Fig. 2A are results generated by an approximate Bayesian approach, which has been established as the currently best-performing algorithm in identifying contacts from sequence correlation analysis (10). The Bayesian approach (yellow ▼ symbol) is seen to perform better than the simple co-variance analysis (MI), but TP rates are not as high as the ones obtained by mfDCA. Analogous results for the relative performance of these methods are also observed for a collection of 25 eukaryotic proteins analyzed (see Fig. S3), suggesting that the applicability of DCA is not restricted to bacterial proteins.

As seen in Fig. 2A, on average 84% of the top-20 DI pairs found by mfDCA (✶ symbol, black solid curve), are true contacts. The average TP rate is indicative of the TP of typical domain families, as the individual TPs for the 131 families examined are distributed mostly in the range of 0.7-1.0; see Fig. S4A evaluated using the best-predicted structure and Fig. S4B when all 856 structures are used. Fig. S4 also shows little difference in the quality of the prediction using the top-10, 20, or 30 DI pairs, and coherent results between the best-predicted and all 856 structures, despite the somewhat uneven distribution of available PDB structures over the 131 domain families. The distribution of the actual (minimum atomic) intra-domain distances between residue pairs with the top 10, 20 and 30 DI ranking are shown in Fig. 2B, using the complete set of 856 PDB structures. The distribution exhibits a strong peak around 3-5Å with a weaker secondary peak around 7-8Å, for all 3 sets of DI rankings used. This double-peak structure is a characteristic feature of the DCA results. It is not observed in the background distribution of all residue pairs, cf. Fig. S5, which has a single maximum around 20-25Å. In Fig. 2B, this background is reflected by a small bump in the histograms for the top 20 and 30 DI ranking pairs. The two short-distance peaks are consistent with the biophysics of molecular contacts: The first peak presumably arises from short-ranged interactions like hydrogen bonding or pairings involved in secondary structure formation, while the second peak likely corresponds to long-ranged, possibly water mediated, contacts (28-30). The observation of this second, biologically reasonable peak in Fig. 2B also motivates the choice of 8Å as a cutoff distance for what is considered a residue-residue contacts in Figs. 1 and 2A.

---

[†] The best-predicted structures were used due to the variance in the quality of PDB structures. Also, for the number of cases where substantially different structures of the same protein exists in the PDB, the existence of a single structure containing the predicted contacts substantiates them as contacts of a native conformation of that protein.

[‡] Both DI and MI benefited modestly from sampling correction; see Fig. S1 for a comparison of the performance of these methods with/without sampling correction.

To understand how many sequences are actually needed for mfDCA, we randomly generated sub-alignments for two protein families; see Fig. S6. At least for these two families, an effective number of $M_{eff} \approx 250$ is already sufficient to reach TP rates close to one for the top predicted residue pairs, and the predictive power is increasing monotonously when more sequences are available. These numbers are consistent with but slightly larger than the sequence requirements reported in Ref. (31) for the statistical-coupling analysis originally proposed in (5).

**Long-distance high-DI residue pairs**

The results from the previous section illustrate the ability of mfDCA to identify intra-domain contacts with high sensitivity. However, a small fraction of pairs showed high DI values (in the top 20-30 ranking) but were located far away according to the available crystal structure. Here we investigate various biological reasons for the appearance of such long distance direct correlations.

**Inter-domain residue contacts.** Given the biological role of some inter-domain contacts (32) we studied if the appearance of long-distance high-DI pairs may be due to interactions between proteins which form oligomeric complexes, as described previously for the dimeric response regulators of the bacterial two-component signaling system (16). To further investigate this possibility, we examined members of the 131 proteins which formed homo-dimers or higher-order oligomers according to the corresponding X-ray crystal structures.

A first example is the ATPase domain of the family of the NtrC-like sigma54-dependent transcriptional activators (Pfam PF00158). Upon activation, different subunits of this domain are known to pack in the front-to-back orientation to form a heptameric ring, wrapping DNA around the complex (33). We compared the DCA results to the structure of NtrC1 of *A. aeolicus* (PDB 1ny6 (33)). Among the top-20 DI pairs, 17 were intra-domain contacts. The 3 remaining pairs were long-distance (> 10 Å) within the domain. Strikingly, all three were within 5 Å when paired with the closest position in an adjacent subunit of the heptamer complex; see Fig. 3. These pairs appear to have co-evolved to maintain the proper formation of the heptamer complex. A second example of high-DI inter-domain contact is shown in Fig. S7 for the protein MexA of *P. aeruginosa*, where 9 subunits oligomerize to form a funnel-like structure across the periplasmic space for antibiotic efflux (PDB 1vf7 (34)).

We further tested the occurrence of inter-domain contacts at a global level. Out of the 131 studied domain families, 21 families feature X-ray crystal structures involving oligomers with predicted inter-domain contacts (see Table S3). Among the top-20 DI pairs which are not intra-domain contacts, about half of them turned out to be inter-domain contacts as shown in Fig. 3D.

**Alternative domain conformations.** Another cause of long-distance high-DI pairs is the occurrence of alternative conformations for domains within the same family. As an illustration, we examine the domain family GerE (Pfam PF00196), whose members include the DNA binding domains of many response regulators in two-component signaling systems.

Using the DNA-bound DNA binding domain of the response regulator NarL of *E. coli* (PDB 1je8 (35)) as a structural template, we found that all of the top-20 DI pairs are true contacts (red bonds in Fig. 4A). However, when mapping the same DI pairs to the structure of the full-length response regulator DosR of *M. tuberculosis* (PDB 3c3w (36)), 7 pairs are found at distances > 8Å (green bonds in Fig. 4B, with the response-regulator domain shown in grey). Comparison of Fig. 4A and 4B shows clearly that all of the green bonds involve pairing with the C-terminal helix (shown in light blue), which is significantly displaced in the full-length structure, presumably due to interaction with the (unphosphorylated) regulatory domain. As proposed by Wisedchaisri et al., a likely scenario is that the DNA-binding domain of DosR is broken up by the inter-domain interaction in the absence of phosphorylation, whereas phosphorylation of DosR restores its DNA-binding domain into the active form represented by the DNA-bound NarL structure.

It is difficult to estimate the extent to which alternative conformations may be responsible for the observed long-distance high-DI contacts, for less characterized domains for which alternative conformations may not be known. However, the example shown in Fig. 4 may motivate future studies to use these long-distance high-DI contacts to explore possible alternative conformations.

**Ligand-mediated interactions.** Another special case of inter-domain residue interactions and another cause of long-distance high-DI pairing is shown in Fig. 5. Here, mfDCA found the metalloenzyme domain family (PF00903) to have a high-DI intra-domain residue pair which is separated by more than 14Å when mapped to the enzyme FosA of *P. aeruginosa* (PDB ID 1nki (37)). FosA is a metalloglutathione transferase which confers resistance to fosfomycin by catalyzing the addition of glutathione to fosfomycin. It is a homodimeric enzyme whose activity is dependent on Mn(II) and $K^+$, and the Mn(II) center has been proposed as part of the catalytic mechanism (37). We observed that the two residues belonging to the different subunits of the high-DI pair, Glu110 (pink) and His7 (yellow), are in direct contact (3 Å residue pair and 1.5 Å residue-ligand separation) with the Mn(II) ion (red) in the dimer configuration (Fig. 5). Thus, the "direct interaction" between these residues found by mfDCA is presumably mediated through their common interaction with a third agent, the metal ion in this case. There may well be other cases with interactions mediated by binding to other metabolites, RNA, DNA or proteins not captured in the available crystal structures.

**Contact map reconstruction**

So far, we have focused on the top-20 DI pairs, which are largely intra- or inter-domain contacts. However, one of the most striking features of the DI result in Fig. 2A is how gradually the average TP rate declines with increasing DI ranking. It is therefore possible to turn the question around: How many residue pairs are predicted, when we require a given minimum TP rate? For instance, one can go up to a DI-ranking of 70 before the average TP rate declines to 70%, meaning that if one were to predict contacts using the top 70 DI pairs, one would have obtained ~50 true contacts on average. This feature may be exploited for sequence-based structure prediction and deserves further analysis.

To become more quantitative, we define the *Number of Acceptable Pairs* $NAP_x$ as the (largest) number of DI-ranked pairs where the specified TP rate (x%) is reached for a given protein. $NAP_x$ can be viewed as an index that characterizes the number of contact predictions at a certain acceptable quality level (given by x). We computed this index for every domain in all 856 structures in our database, for TP levels of 0.9, 0.8 and 0.7. The results are shown as cumulative distributions in Fig. 6. A casual inspection of these distributions shows that there are many structures with high NAP. Suppose the acceptable TP level is 0.7. The median of $NAP_{70}$ is 52, meaning that in half of the structures examined, the number of high-ranking, predictive DI pairs is at least 52. Furthermore, 70% of the structures have $NAP_{70} > 30$ and 34% of the structures have $NAP_{70} > 100$. A normalized version of Fig. 6 with respect to the length of the domain L is shown in Fig. S8. In one extreme case involving the family of bacterial tripartite tricarboxylate receptors (PF03401), $NAP_{70}$ was 600, i.e., 70% of the top 600 DI pairs correspond to true contacts when mapped to the best-predicted structure (PDB ID 2qpq (38)), see Fig. S9A. This domain has a length of L=274 and has ~2300 contacts. In another example, the extracellular solute-binding family (PF00496) mapped to the structure of the periplasmic oligopeptide-binding protein OppA of *S. typhimurium* (PDB ID 1jet (39)) has a $NAP_{70}$ of 497 (Fig. S9B, L=372 and ~2530 contacts).

We computed also the $NAP_{70}$ distribution using MI; see Fig. S10. The difference between DI and MI, about 10-20% in TP rate according to Fig. 2A, is seen much more significantly when displayed according to the NAP index, with the median $NAP_{70}$ being 5 for MI and 52 for DI. This shows that DCA generates many more high-valued contact pair predictions. We also compared the performance of DCA with the approximate Bayesian method (red dashed curve in Fig. S10), which gives a median $NAP_{70}$ of 25 that is halfway between that of MI and mfDCA.

The large number of contacts correctly predicted by DCA prompted us to explore the extent to which DCA may be used to predict the contact maps of protein domains. For a domain with *L* amino acids, we calculated the inferred maps by sorting residue pairs according to their DIs, and keeping the *2L* highest-ranking pairs with minimum separation of 5 positions along the sequence. For the contact map prediction we included further those pairings which have equal or larger DI than the ones mentioned above, but with shorter separation along the sequence as they may be informative about secondary structures. Fig. 7 shows two examples of such contact map predictions, for the prokaryotic promoter recognition domain of SigmaE already shown in Fig. 1 (PDB 1or7, *L*=71) and for the eukaryotic signaling protein

Ras (PDB 5p21 (40), *L*=160). The figure shows the native contact maps, together with the predictions by MI (left panels) and DI (central panels). Correctly predicted native contacts (i.e., the TPs) are indicated in red. The unpredicted native contacts taken from the X-ray crystal structures are shown in grey, and the incorrect predictions in green. It is evident that for both proteins, DI works substantially better than MI, both in terms of the TP rate and the representation of the native contact map. To become more quantitative, we have binned the predicted pairs according to their separation along the primary amino-acid sequence (right panels in Fig. 7). We observe that DI captures in particular a higher number and more accurately those contacts between residues, which are very distant along the sequence. Also, the DI predictions are more evenly distributed, whereas MI predictions tend to cluster together.

## Conclusion and Perspectives

We have shown the ability of DCA to identify with high accuracy residue pairs in domain families that might have co-evolved together and hence are representative of physical proximity in the three-dimensional fold of the domain. We have done an extensive evaluation of these capabilities for a large number of families and individual PDB structures. We found that DCA is not only able to identify intra-domain contacts but also inter-domain residue pairs that are part of oligomerization interfaces. Although we focused on bacterial proteins, this methodology can be applied to the ever-increasing number of eukaryotic sequences. Our initial results suggest that mfDCA performance is conserved for non-bacterial proteins. One potential application is the identification of interaction interfaces for homo-dimers that could ultimately help in complex structure prediction, e.g. the cases in Fig. 3 and S7. Our results might open new avenues of research for which full contact maps could be estimated and used as input data for *de novo* protein structure identification, which is particularly interesting in the case of inter-domain contacts in multi-domain proteins (in preparation). Ultimately, this methodology can be utilized with pairs of proteins rather than single proteins to identify potential protein-protein interactions. An example of this approach was introduced in (16), however, the current mathematical formulation of the method as well as its computational implementation allows an analysis to a much larger scale.

Despite the accuracy of the extracted signal, mfDCA cannot be expected to extract all biological information contained in the pair correlations. This can be illustrated by comparing the mfDCA results to those of Statistical Coupling Analysis (SCA), developed by Ranganathan and coworkers (5) and used to identify "co-evolving protein sectors" (41). We have applied mfDCA to the data of (41) for the Trypsin protein family (Serine protease), where SCA identified 3 sectors related to different functionalities of the protein, which cover almost 30% of all residues. mfDCA leads to a 83.3% TP rate for the top-30 contact predictions (PDB 3tgi (42)), i.e. to a performance which is comparable to the other protein families analyzed here. Out of the resulting 25 true contact pairs, only 8 are found within the identified sectors. Among them, three are disulfide bonds (C42:C58, C136:C201, C191:C220), and another two are inside a catalytic triad crucial for the catalytic activity of the protein family (H57:S195, D102:S195). The other 17 true contacts predicted by mfDCA are distributed over the protein fold, without obvious relation to the sectors (See Table S4). The difference in prediction can be traced back to differences in the algorithmic approaches: SCA uses clustering to identify larger groups of co-evolving sites (sectors), whereas DCA uses maximum-entropy modeling to extract pairs of directly coupled residues. Thus, the two algorithms extract different and, in both cases, biologically important information. It remains a future challenge to develop techniques unifying SCA and DCA, and to extract even more co-evolutionary information from multiple-sequence alignments.

## Methods

**Data extraction**

Sequence data sets were extracted primarily from Pfam families with more than 1000 non-redundant sequences. We decided to focus on families that are predominantly bacterial (i.e. more than 90% of the family sequences belong to bacterial organisms). Another requirement in this dataset is that such families must have at least two known X-ray crystal structures with a resolution of 3Å or better. The Protein Data Bank (PDB) (43) was accessed to obtain crystal structures of proteins. An additional criterion to improve statistical significance when picking sequences that belong to a particular Pfam (22) family, was to use a stricter E-value threshold than the standard used by HMMer (21) to classify domain membership. An in-house mapping application was developed to map domain family alignments and predicted couplets to specific residues in PDB structures. Some of the data extraction tools used in this study are described in more detail in (17). A total of 131 families were selected which complied with all these criteria. A list of these Pfam families and the 856 PDB structures analyzed can be accessed in Supplementary information (Table S1-S2).

For each family, the protein sequences are collected in one MSA denoted by $\{(A_1^a,...,A_L^a) \mid a=1,...,M\}$, where $L$ denotes the number of MSA columns, i.e. the length of the protein domains. Alignments are local alignments to the Pfam Hidden Markov Models, due to the large number of proteins in each MSA we refrained from refinements using global alignment techniques.

**Sequence statistics and reweighting**

As already mentioned in Results and Discussion, the main inputs of DCA are reweighted frequency counts for single MSA columns and column pairs:

$$f_i(A) = \frac{1}{M_{eff} + \lambda}\left(\frac{\lambda}{q} + \sum_{a=1}^{M}\frac{1}{m^a}\delta_{A,A_i^a}\right)$$

$$f_{ij}(A,B) = \frac{1}{M_{eff} + \lambda}\left(\frac{\lambda}{q^2} + \sum_{a=1}^{M}\frac{1}{m^a}\delta_{A,A_i^a}\delta_{B,A_j^a}\right)$$

(1)

In this equation, $\delta_{A,B}$ denotes the Kronecker symbol, which equals one if $A = B$, and zero otherwise. Further more, we have defined $q = 21$ for the number of different amino acids (counting also the gap), and a pseudo-count $\lambda$ (44), whose value will be discussed below. The weighting of the influence of a single sequence by the factor $1/m^a$ aims at correcting for the sampling bias. It is determined by the number

$$m^a = \left|\left\{b \in \{1,...,M\} \mid seqid(A^a, A^b) > 80\%\right\}\right|$$

(2)

of sequences $A^b = (A_1^b,...,A_L^b), b \in \{1,...,M\}$, which have more than 80% sequence identity ($seqid$) with $A^a = (A_1^a,...,A_L^a)$, where $a$ itself is counted. The same reweighting but with a 100% sequence-identity threshold would remove multiple counts of repeated sequences. Reweighting systematically improves the results, see Fig. S1, with only a weak dependence on the precise threshold value (in the range 70-90%) and the specific protein family. Last, we introduced the effective sequence number $M_{eff} = \sum_{a=1}^{M} 1/m^a$ as the sum over all sequence weights.

These counts allow for calculating the mutual information,

$$MI_{ij} = \sum_{A,B} f_{ij}(A,B) \ln \frac{f_{ij}(A,B)}{f_i(A)f_j(B)} , \qquad (3)$$

which equals zero if and only if $i$ and $j$ are uncorrelated, and is positive else.

**Maximum-entropy modeling**

To disentangle direct and indirect couplings, we aim at inferring a statistical model $P(A_1,...,A_L)$ for entire protein sequences $(A_1,...,A_L)$. To achieve coherence with data, we require this model to generate the empirical frequency counts as marginals,

$$\begin{aligned}\forall i, A_i : & \sum_{\{A_k | k \neq i\}} P(A_1,...,A_L) \equiv f_i(A_i) \\ \forall i,j, A_i, A_j : & \sum_{\{A_k | k \neq i,j\}} P(A_1,...,A_L) \equiv f_{ij}(A_i,A_j)\end{aligned} \qquad (4)$$

Besides this constraint, we aim at the most general, least constrained model $P(A_1,...,A_L)$. This can be achieved by applying the *maximum-entropy principle* (45, 46), and leads to an explicit mathematical form of $P(A_1,...,A_L)$ as a Boltzmann distribution with pairwise couplings $e_{ij}(A,B)$ and local biases (fields) $h_i(A)$:

$$P(A_1,...,A_L) = \frac{1}{Z} \exp\left\{ \sum_{i<j} e_{ij}(A_i,A_j) + \sum_i h_i(A_i) \right\} . \qquad (5)$$

The model parameters have to be fitted such that Eq. (4) is satisfied. In this fitting procedure, one has to consider that Eq. (5) contains more free parameters than there are independent conditions in Eq. (4), which allows to change couplings and fields together without changing the sum in the exponent. Therefore, multiple but equivalent solutions for the fitting are possible. To remove this freedom, we consider all couplings and fields measured relative to the last amino acid $A = q$, and set

$$\forall i,j,A : \quad e_{ij}(A,q) = e_{ij}(q,A) = 0, \quad h_i(q) = 0 , \qquad (6)$$

Details on the maximum-entropy approach are given in the Supplement.

**Small-coupling expansion**

Eq. (5) contains the normalization factor $Z$, in statistical physics also called the *partition function*, which is defined as

$$Z = \sum_{A_1,...,A_L} \exp\left\{\sum_{i<j} e_{ij}(A_i,A_j) + \sum_i h_i(A_i)\right\} \tag{7}$$

and includes a sum of $q^L$ terms. Its direct calculation is infeasible for any realistic protein length, and approximations have to be used. In a prior paper (16), several of us introduced a message-passing approach, which allows the treatment of about 70 MSA columns simultaneously in about two days running time on a standard desktop PC (larger MSAs need preprocessing to decrease the number of columns before running message passing). Here we introduce a much more efficient scheme, which for $L = 70$ is about 3-4 orders of magnitude faster, and which allows to directly analyze alignments with $L \leq 1000$ ($L \leq 500$ on a standard PC due to limited working memory). The total algorithmic complexity is $O(q^3 N^3)$. The major speedup compared to the iterative message-passing solver results from the fact that parameter inference can be done in a single computational step in the new algorithm.

The approach is based on a small-coupling expansion (47, 48), which is explained in detail in the Supplement: The exponential of $\Sigma_{i<j} e_{ij}(A_i,A_j)$ in Eq. (7) is expanded into a Taylor series. Keeping only the linear order of this expansion, we obtain the well-known mean-field equations

$$\frac{f_i(A)}{f_i(q)} = \exp\left\{h_i(A) + \sum_A \sum_{j \neq i} e_{ij}(A,B) f_j(B)\right\}, \tag{8}$$

containing the single-column counts, as well as a simple relation between the coupling $e_{ij}(A,B)$ and the pair counts $f_{ij}(A,B)$ for all $i,j = 1,...,L$ and $A,B = 1,...,q-1$

$$e_{ij}(A,B) = -\left(C^{-1}\right)_{ij}(A,B) \tag{9}$$

where

$$C_{ij}(A,B) = f_{ij}(A,B) - f_i(A) f_j(B). \tag{10}$$

Eqs. (6) and (9) completely determine the couplings in terms of the data. Note that the connected-correlation matrix $C$ defined in Eq. (10) is a $(q-1)L \times (q-1)L$ matrix, the pairs $(i,A)$ and $(j,B)$ have to be understood as joint single indices in the inversion in Eq. (9).

In general, when constructed without pseudo-counts ($\lambda = 0$), this matrix is not invertible, and formally Eq. (9) leads to infinite couplings. Even introducing site-specific reduced amino-acid alphabets (only those actually observed in the corresponding MSA column) is found to be not sufficient for invertibility. The matrix can, however, be regularized by setting $\lambda > 0$. For small $\lambda$, elements diverging in the $\lambda \to 0$ limit dominate the DI calculation discussed in the next paragraph. To avoid this, we have to go to relatively large pseudo-counts, $\lambda = M_{eff}$ is found to be a reasonable value throughout families, and is used exclusively in this paper. Fig. S11 shows a sensitivity analysis for different values of the pseudo-count for two domain families. The mean TP rates are computed for pseudo-count values $\lambda = w \cdot M_{eff}$, with the weights $w$ ranging from 0.11 to 9. The optimum value of $\lambda$ is found for $1 \leq w \leq 1.5$. Therefore, we used $\lambda = M_{eff}$ throughout this study.

Due to the long run time of the message-passing approach (mpDCA), we could not compare its performance for all proteins studied in this paper. Fig. S12 contains two examples, Trypsin (PF00089) and Trypsin inhibitor (PF00014). In both cases, mfDCA outperforms mpDCA. Furthermore, it is straightforward to include into DCA also the next order of the small-coupling expansion beyond the mean-field approximation (This corresponds to the so-called TAP equations in spin-glass physics (49)). We do not find any systematic improvement of the resulting algorithm, called tapDCA, when compared to mfDCA; see Fig. S12.

**Direct information**

After having estimated the direct coupling $e_{ij}(A,B)$ through Eq. (8), we need a strategy for ranking the $L(L-1)$ possible interactions according to their *direct coupling* strength. Following the idea that MI is a good measure for correlations, in (16) we introduced a quantity called *direct information* (DI). It can be understood as the amount of MI between columns $i$ and $j$, which results from direct coupling alone.

To this end, we introduce for each column pair $(i,j)$ an isolated two-site model

$$P_{ij}^{(dir)}(A,B) = \frac{1}{Z_{ij}} \exp\{e_{ij}(A,B) + \tilde{h}_i(A) + \tilde{h}_j(B)\}, \qquad (11)$$

where the couplings $e_{ij}(A,B)$ are computed using Eq. (8), and the auxiliary fields $\tilde{h}$ are given implicitly by compatibility with the empirical single-residue counts:

$$f_i(A) = \sum_B P_{ij}^{(dir)}(A,B), \qquad f_j(B) = \sum_A P_{ij}^{(dir)}(A,B) \qquad (12)$$

As before, in order to reduce the number of free parameters to the number of independent constraints, these fields are required to fulfill $\tilde{h}_i(q) = \tilde{h}_j(q) = 0$. Note that the auxiliary fields have to be determined for each pair $(i,j)$ independently to fit Eq. (12). Finally, we define the direct information DI as the MI of model (11),

$$DI_{ij} = \sum_{AB} P_{ij}^{(dir)}(A,B) \ln \frac{P_{ij}^{(dir)}(A,B)}{f_i(A) f_j(B)} \qquad (13)$$

**Algorithmic implementation**

The algorithmic implementation of the mean-field approximation is sketched in the following steps:

1. Estimate the frequency counts $f_i(A)$ and $f_{ij}(A,B)$ from the MSA, using the pseudo-count $\lambda = M_{eff}$ in Eq. (1) and (2).
2. Determine the empirical estimate of the connected correlation matrix Eq. (10).
3. Determine the couplings $e_{ij}(A,B)$ according to the second of Eqs. (9).
4. For each column pair $i < j$, estimate the direct information $DI_{ij}$ by solving Eq. (11-12) for $P_{ij}^{(dir)}(A,B)$, and plug the result into Eq. (13).

An implementation of the code in Matlab is available upon request.

## Acknowledgments

We thank Hendrik Szurmant, Joanna Sulkowska and Lucy Colwell for useful discussions during the course of this work. This work was supported by the Center for Theoretical Biological Physics sponsored by the NSF (Grant PHY-0822283) and by NSF- MCB-1051438 through JNO.

# References

1.  Altschuh D, Lesk A, Bloomer A, Klug A (1987) Correlation of co-ordinated amino acid substitutions with function in viruses related to tobacco mosaic virus. *J Mol Bio* 193:693–707.
2.  Göbel U, Sander C, Schneider R, Valencia A (1994) Correlated mutations and residue contacts in proteins. *Proteins: Structure, Function, and Genetics* 18:309–317.
3.  Neher E (1994) How frequent are correlated changes in families of protein sequences? *Proc Natl Acad Sci USA* 91:98-102.
4.  Shindyalov IN, Kolchanov NA, Sander C (1994) Can three-dimensional contacts in protein structures be predicted by analysis of correlated mutations? *Protein Engineering* 7:349-358.
5.  Lockless SW, Ranganathan R (1999) Evolutionarily conserved pathways of energetic connectivity in protein families. *Science* 286:295-299.
6.  Atchley WR, Wollenberg KR, Fitch WM, Terhalle W, Dress AW (2000) Correlations among amino acid sites in bHLH protein domains: an information theoretic analysis. *Mol Biol Evol* 17:164-178.
7.  Fodor AA, Aldrich RW (2004) Influence of conservation on calculations of amino acid covariance in multiple sequence alignments. *Proteins: Structure, Function, and Bioinformatics* 56:211–221.
8.  Liu Z, Chen J, Thirumalai D (2009) On the accuracy of inferring energetic coupling between distant sites in protein families from evolutionary imprints: Illustrations using lattice model. *Proteins: Structure, Function, and Bioinformatics* 77:823–831.
9.  Lashuel, HA and Pappu R (2009) Amyloids Go Genomic: Insights Regarding the Sequence Determinants of Prion Formation from Genome-Wide Studies. *ChemBioChem* 10:1951-1954.
10. Burger L, Nimwegen E van (2010) Disentangling direct from indirect co-evolution of residues in protein alignments. *PLoS Computational Biology* 6:e1000633.
11. Liolios K et al. (2010) The Genomes On Line Database (GOLD) in 2009: status of genomic and metagenomic projects and their associated metadata. *Nucleic Acids Research* 38:D346-D354.
12. Hoch JA (2000) Two-component and phosphorelay signal-transduction. *Current Opinion in Microbiology* 3:165–170.
13. Ulrich LE, Zhulin IB (2010) The MiST2 database: a comprehensive genomics resource on microbial signal transduction. *Nucleic Acids Research* 38:D401-D407.
14. White RA, Szurmant H, Hoch JA, Hwa T (2007) Features of Protein–Protein Interactions in Two-Component Signaling Deduced from Genomic Libraries. *Methods in Enzymology* 422:75-101.
15. Skerker JM et al. (2008) Rewiring the specificity of two-component signal transduction systems. *Cell* 133:1043-1054.
16. Weigt M, White RA, Szurmant H, Hoch JA, Hwa T (2009) Identification of direct residue contacts in protein-protein interaction by message passing. *Proc Natl Acad Sci USA* 106:67-72.
17. Lunt B et al. (2010) Inference of direct residue contacts in two-component signaling. *Methods in Enzymology* 471:17–41.
18. Schug A, Weigt M, Onuchic JN, Hwa T, Szurmant H (2009) High-resolution protein complexes from integrating genomic information with molecular simulation. *Proc Natl Acad Sci USA* 106:22124-22129.
19. Schug A, Weigt M, Hoch J, Onuchic J (2010) Computational modeling of phosphotransfer complexes in two-component signaling. *Methods in Enzymology* 471:43-58.
20. Procaccini A, Lunt B, Szurmant H, Hwa T, Weigt M (2011) Dissecting the specificity of protein-protein interaction in bacterial two-component signaling: orphans and crosstalks. *PLoS one* 6:e19729.
21. Eddy SR (1998) Profile hidden Markov models. *Bioinformatics* 14:755-763.
22. Finn RD et al. (2010) The Pfam protein families database. *Nucleic Acids Research* 38:D211-D222.
23. Wollenberg KR, Atchley WR (2000) Separation of phylogenetic and functional associations in biological sequences by using the parametric bootstrap. *Proc Natl Acad Sci USA* 97:3288-3291.

24. Tillier ERM, Lui TWH (2003) Using multiple interdependency to separate functional from phylogenetic correlations in protein alignments. *Bioinformatics* 19:750-755.
25. Gouveia-Oliveira R, Pedersen AG (2007) Finding coevolving amino acid residues using row and column weighting of mutual information and multi-dimensional amino acid representation. *Algorithms for Molecular Biology* 2:1-12.
26. Dunn SD, Wahl LM, Gloor GB (2008) Mutual information without the influence of phylogeny or entropy dramatically improves residue contact prediction. *Bioinformatics* 24:333-340.
27. Campbell E a et al. (2003) Crystal structure of Escherichia coli sigmaE with the cytoplasmic domain of its anti-sigma RseA. *Molecular Cell* 11:1067-1078.
28. Tanaka S, Scheraga H A (1976) Medium- and long-range interaction parameters between amino acids for predicting three-dimensional structures of proteins. *Macromolecules* 9:945-950.
29. Go N, Taketomi H (1978) Respective roles of short- and long-range interactions in protein folding. *Proc Natl Acad Sci USA* 75:559-563.
30. Miyazawa S, Jernigan RL (2003) Long- and short-range interactions in native protein structures are consistent/minimally frustrated in sequence space. *Proteins: Structure, Function, and Genetics* 50:35-43.
31. Dima, RI and Thirumalai D (2006) Determination of network of residues that regulate allostery in protein families using sequence analysis. *Prot. Sci.* 15:258-268.
32. Myers RS, Amaro RE, Luthey-Schulten Z a, Davisson VJ (2005) Reaction coupling through interdomain contacts in imidazole glycerol phosphate synthase. *Biochemistry* 44:11974-11985.
33. Lee S-Y et al. (2003) Regulation of the transcriptional activator NtrC1: structural studies of the regulatory and AAA+ ATPase domains. *Genes Dev* 17:2552-2563.
34. Akama H et al. (2004) Crystal structure of the membrane fusion protein, MexA, of the multidrug transporter in Pseudomonas aeruginosa. *J Biol Chem* 279:25939-25942.
35. Maris AE et al. (2002) Dimerization allows DNA target site recognition by the NarL response regulator. *Nat Struc Biol* 9:771-778.
36. Wisedchaisri G, Wu M, Sherman DR, Hol WGJ (2008) Crystal structures of the response regulator DosR from Mycobacterium tuberculosis suggest a helix rearrangement mechanism for phosphorylation activation. *J Mol Biol* 378:227-242.
37. Rigsby RE, Rife CL, Fillgrove KL, Newcomer ME, Armstrong RN (2004) Phosphonoformate: a minimal transition state analogue inhibitor of the fosfomycin resistance protein, FosA. *Biochemistry* 43:13666-13673.
38. Herrou J et al. (2007) Structure-based mechanism of ligand binding for periplasmic solute-binding protein of the Bug family. *J Mol Biol* 373:954-964.
39. Tame JRH, Sleigh SH, Wilkinson AJ, Ladbury JE (1996) the role of water in sequence independent ligand binding by an oligopeptide transporter protein. *Nat Struc Biol* 3:998–1001.
40. Pai EF et al. (1990) Refined crystal structure of the triphosphate conformation of H-ras p21 at 1.35 A resolution: implications for the mechanism of GTP hydrolysis. *The EMBO journal* 9:2351-2359.
41. Halabi N, Rivoire O, Leibler S, Ranganathan R (2009) Protein sectors: evolutionary units of three-dimensional structure. *Cell* 138:774-86.
42. Pasternak A, Ringe D, Hedstrom L (1999) Comparison of anionic and cationic trypsinogens: the anionic activation domain is more flexible in solution and differs in its mode of BPTI binding in the crystal structure. *Protein science* 8:253–258.
43. Berman HM et al. (2000) The Protein Data Bank. *Nucleic Acids Research* 28:235-242.
44. Durbin R, Eddy S, Krogh A, Mitchison G (1998) *Biological sequence analysis: Probabilistic models of proteins and nucleic acids* (Cambridge Univ Pr, New York).
45. Jaynes ET (1957) Information theory and statistical mechanics. *Physical review* 106:620-630.
46. Jaynes ET (1957) Information theory and statistical mechanics. II. *Physical review* 108:171-190.
47. Plefka T (1982) Convergence condition of the TAP equation for the infinite-ranged Ising spin glass model. *J Phys A: Math Gen* 15:1971-1978.
48. Georges A, Yedidia J (1991) How to expand around mean-field theory using high-temperature expansions. *J Phys A Math Gen* 24:2173-2192.
49. Thouless DJ, Anderson PW, Palmer RG (1977) Solution of "Solvable model of a spin glass." *Phil. Mag.* 35:593-601.

# Figure legends

**Figure 1.** Contact predictions for the family of domains homologous to Region 2 of the bacterial Sigma factor (Pfam ID PF04542) mapped to the sequence of the Sigma-E factor of E. coli (encoded by rpoE) (PDB ID 1or7). Panel A shows the top-20 DI predictions and Panel B the top-20 MI predictions for residue-residue contacts, both with a minimum separation of 5 positions along the backbone. Each pair with distance < 8Å is connected by a red link, while the more distant pairs are connected by the green links.

**Figure 2. A)** Mean true positive (TP) rate for 131 domain families, as a function of the number of top-ranked contacts and histogram of the distances of all predicted structures for each of the 131 domains studied. DI results (black ✶ symbol) clearly outperform the other two methods: MI (red O symbol) and an approximate Bayesian approach (yellow ▼ symbol) developed by Burger et al. Burger et al.'s method aims at disentangling direct and indirect correlations by averaging over tree-shaped residue-residue coupling networks, and it contains a phylogeny correction. The method can also reach length-400 multiple alignments as mfDCA does, our implementation follows closely the description in (6). However, coupling trees do not allow for multiple coupling paths between two residues as DCA does, possibly accounting for its lower TP rates compared to mfDCA. **B)** mfDCA predictions for the top 10, 20 and 30 residue pairs show a bimodal distribution of intra-domain distances with two frequency peaks around 3-5Å and 7-8Å.

**Figure 3.** The only 3 long distance high-DI predictions found out of the top-20 DI pairs in the Sigma-54 interaction domain of protein NtrC1 of *A. aeolicus* (PDB: 1ny6) out of the top 20 predicted couplets are multimerization contacts. Structures showing each of these 3 inter-domain contacts which are separated by less than 5 Å in a ring-like heptamer formed by Sigma54 interaction domains. **A)** Residue pair GLU(174)-ARG(253), **B)** residue pair PHE(226)-TYR(261) and **C)** residue pair ALA(197)-ALA(249). **D)** Oligomerization contacts are found in 21 structures of the 131 families studied (see Table S3). These contacts represent a significant percentage of long distance high-DI contacts observed in our predictions.

**Figure 4.** The figures show the top-20 contacts predicted by DI for the family of response regulator DNA binding domain (GerE, PF00196) (containing both the dark and light blue colored regions). In panel A, the contacts are mapped to the DNA binding domain of *E. coli* NarL, bound to the DNA target (PDB 1je8). The TP rate for the top-20 DI pairs is 100%, and are all shown as red links. In panel B, the contacts are mapped to the full-length response regulator DosR of *M. tuberculosis* (PDB 3c3w), with the (unphosphorylated) response regulator domain shown in grey. The top-20 DI pairings is only 65% in this case (13 red and 7 green links). The difference in prediction quality for the two structures can be traced back to a major reorientation of the C-terminal helix of the GerE domain (light blue) in panel B.

**Figure 5.** The metalloenzyme domain (PF00903) of protein FosA (PDB: 1nki) is an example of a case where long-distance high-DI pairs are in fact residue pairs coordinating a ligand. The high-DI pair involving the residues Glu110 (pink) and His7 (yellow) coordinate a metal ion Mn(II) (red) in its dimer configuration. $K^+$ ions are shown as larger spheres (gray and blue), each coordinated by a monomer of the corresponding color.

**Figure 6.** Cumulative distribution of the Number of Acceptable Pairs ($NAP_x$) for a given TP rate x. The curves show the probability of $NAP_x$ to be larger than a given number *n* for contacts at given TP rates of 0.9, 0.8 and 0.7. The curves are computed for all 856 PDB structures in the dataset. We observe that the probability of $NAP_{70} > 30$ is 70% and $NAP_{70} > 100$ is 34%. This implies that a substantial number of protein domains can have accurate predictions that go beyond the top-30 DI pairings. We also identify some exceptional cases with $NAP_{70} > 600$.

**Figure 7.** Two examples of contact map predictions using MI (panel A and D) and mfDCA (panel B and E). Gray symbols represent the native map with a cutoff of 8Å, colored symbols the computational contact predictions using MI or DI ranking (red squares for TP and green squares for spatially distant pairs). The number of pairs is determined such that there are *2L* pairs with minimum separation 5 along the sequence, where *L* is the domain length. The right-most panels (C and F) bin the predictions of MI (blue) and mfDCA (red) according to their separation along the protein sequence. The overall bars count all predictions, the shaded part the TPs. Note in particular that mfDCA leads to a higher number of more accurate predictions for large separations. **A,B,C)** The promoter recognition helix domain of the Sigma-E factor (PDB 1or7). **D,E,F)** The eukaryotic signaling protein Ras (PDB 1p21). For better comparability of native vs. predicted contacts, the predictions are displayed only above the diagonal.

Figure 1

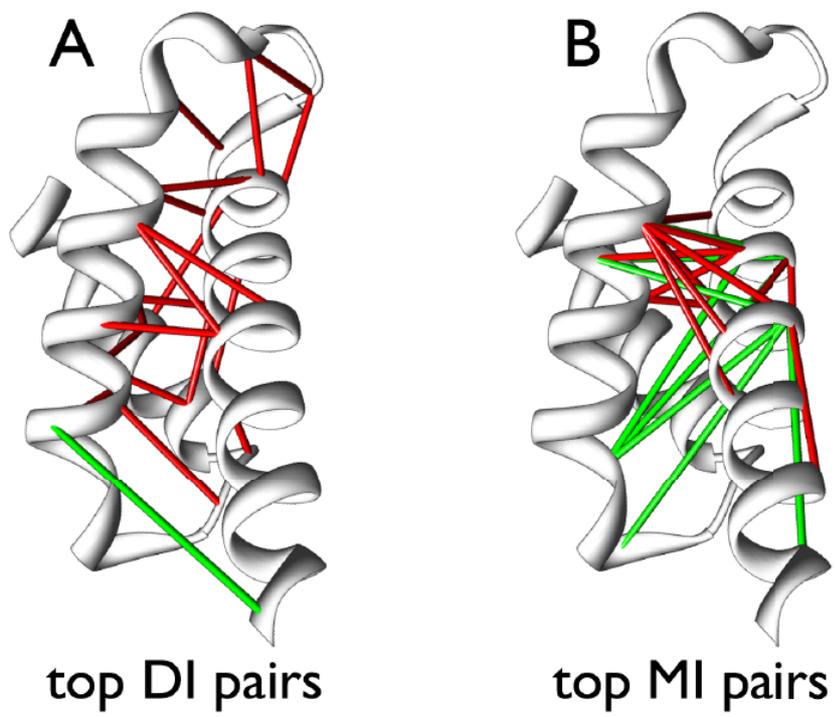

Figure 2

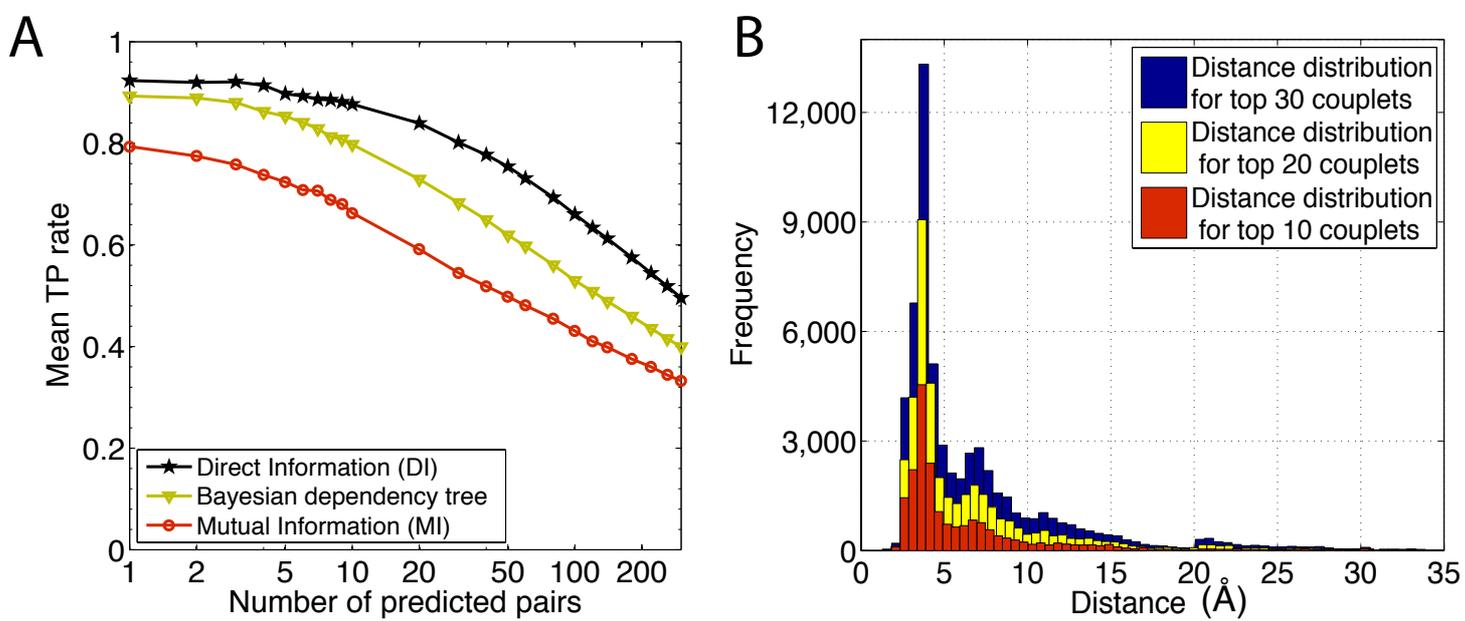

Figure 3

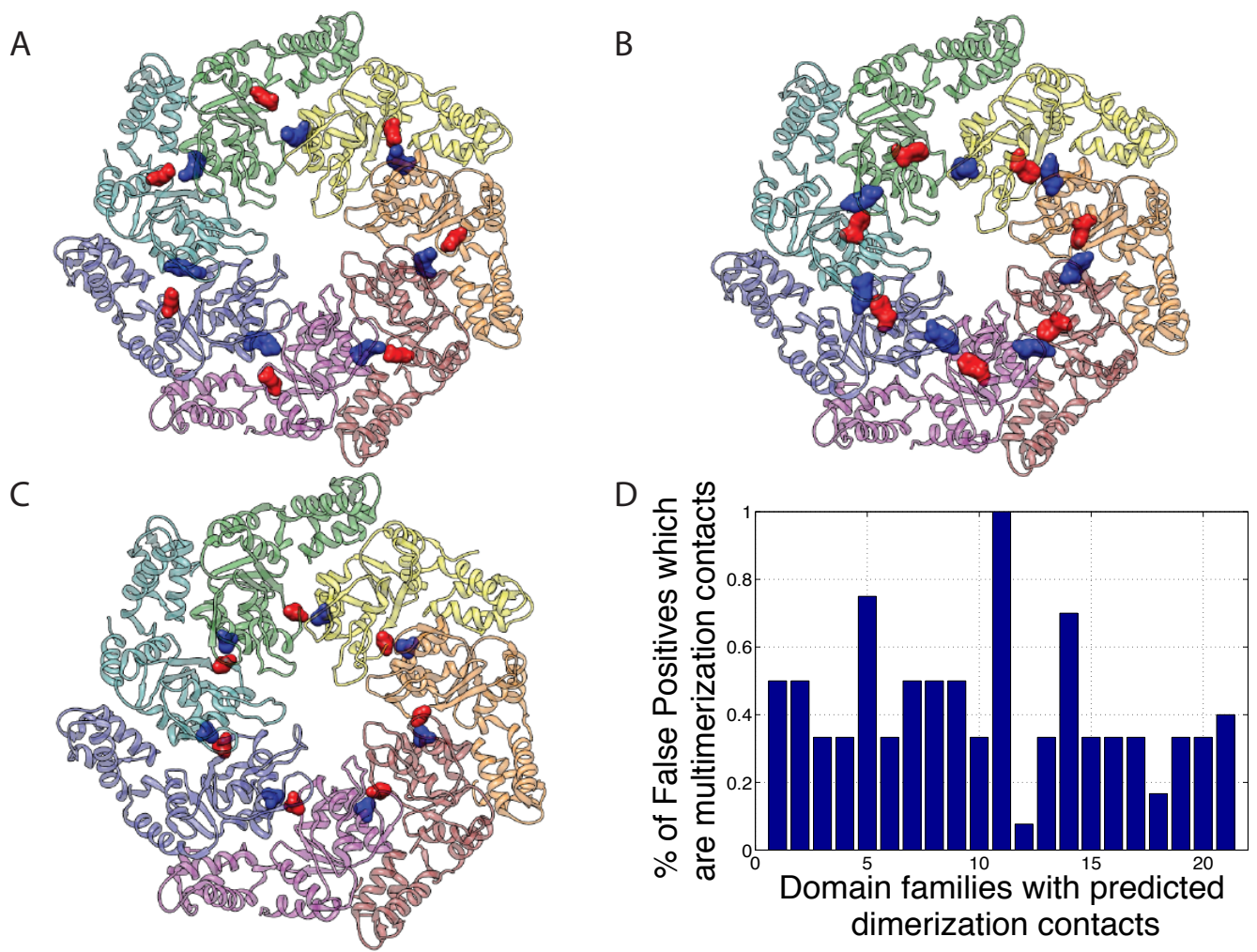

Figure 4

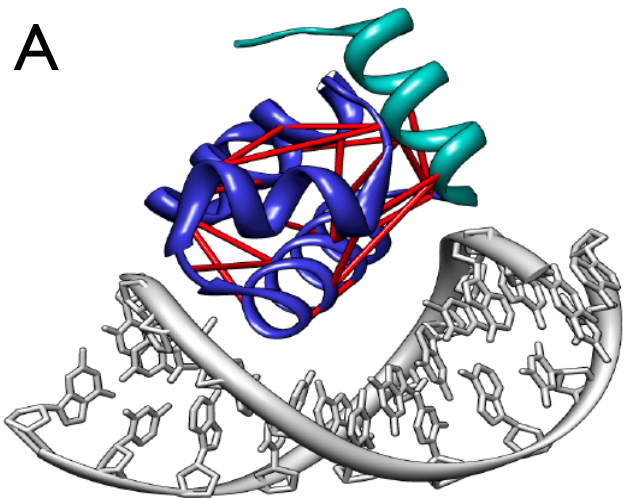 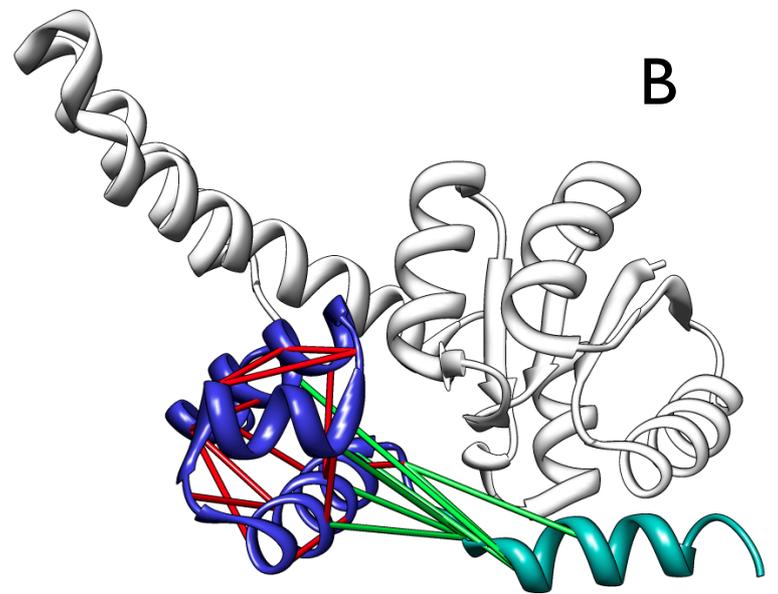

Figure 5

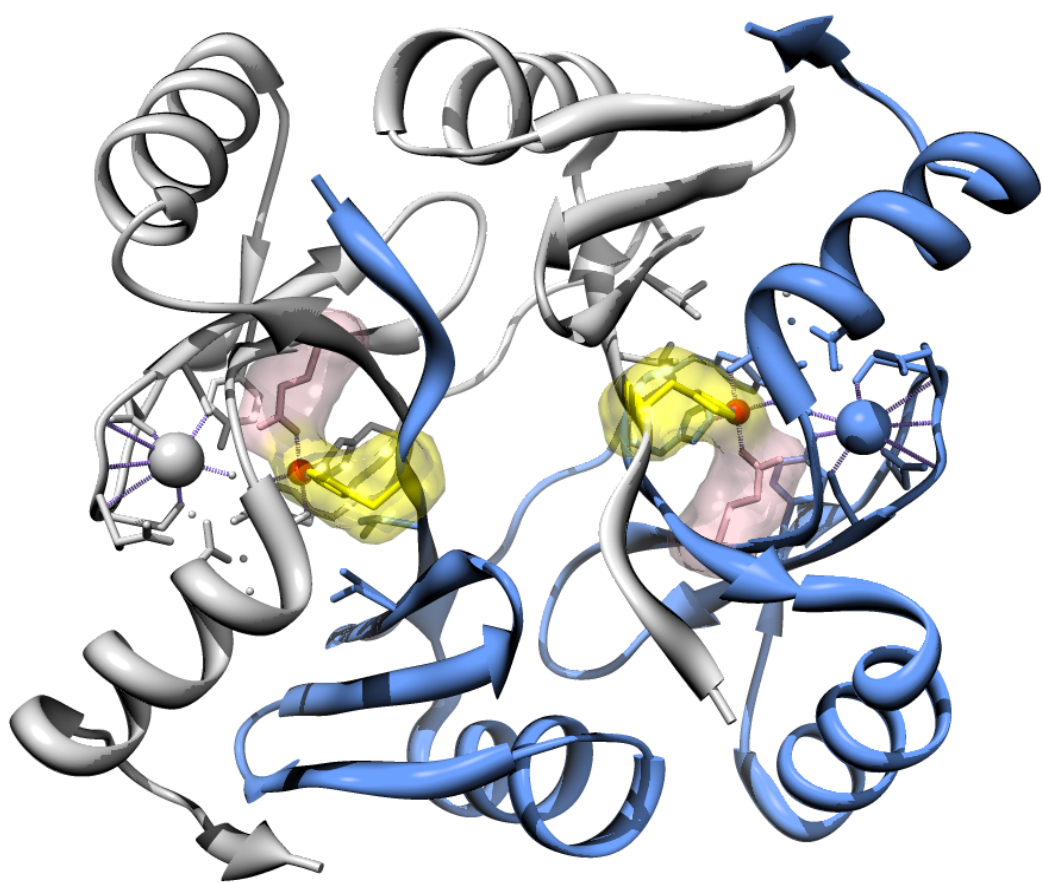

Figure 6

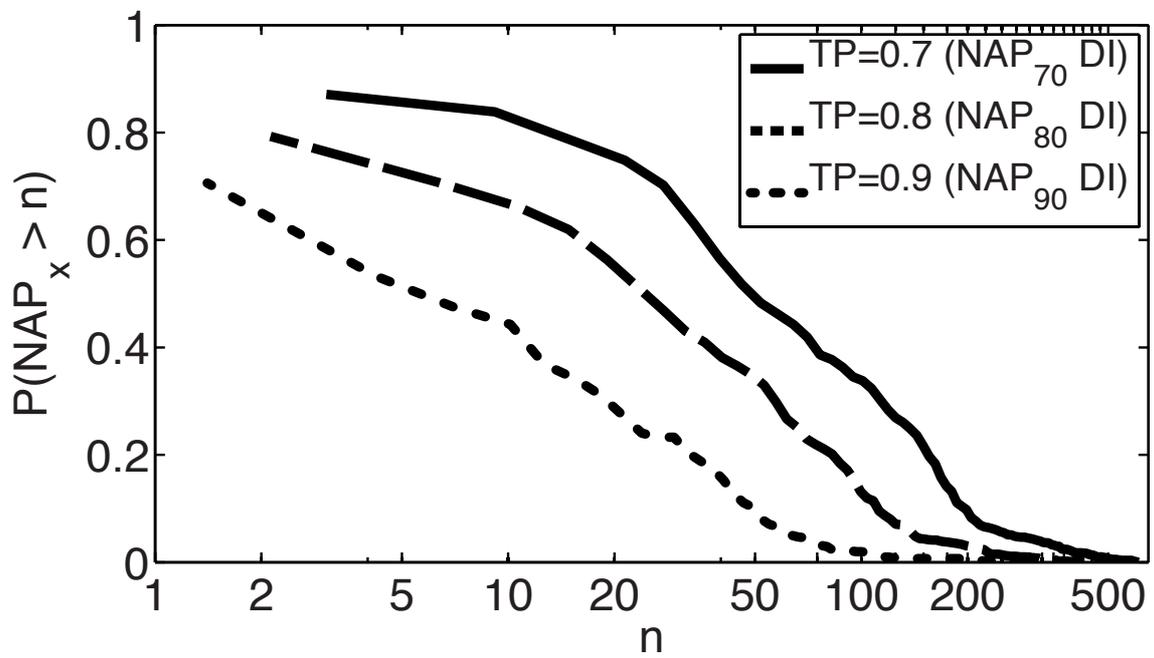

Figure 7

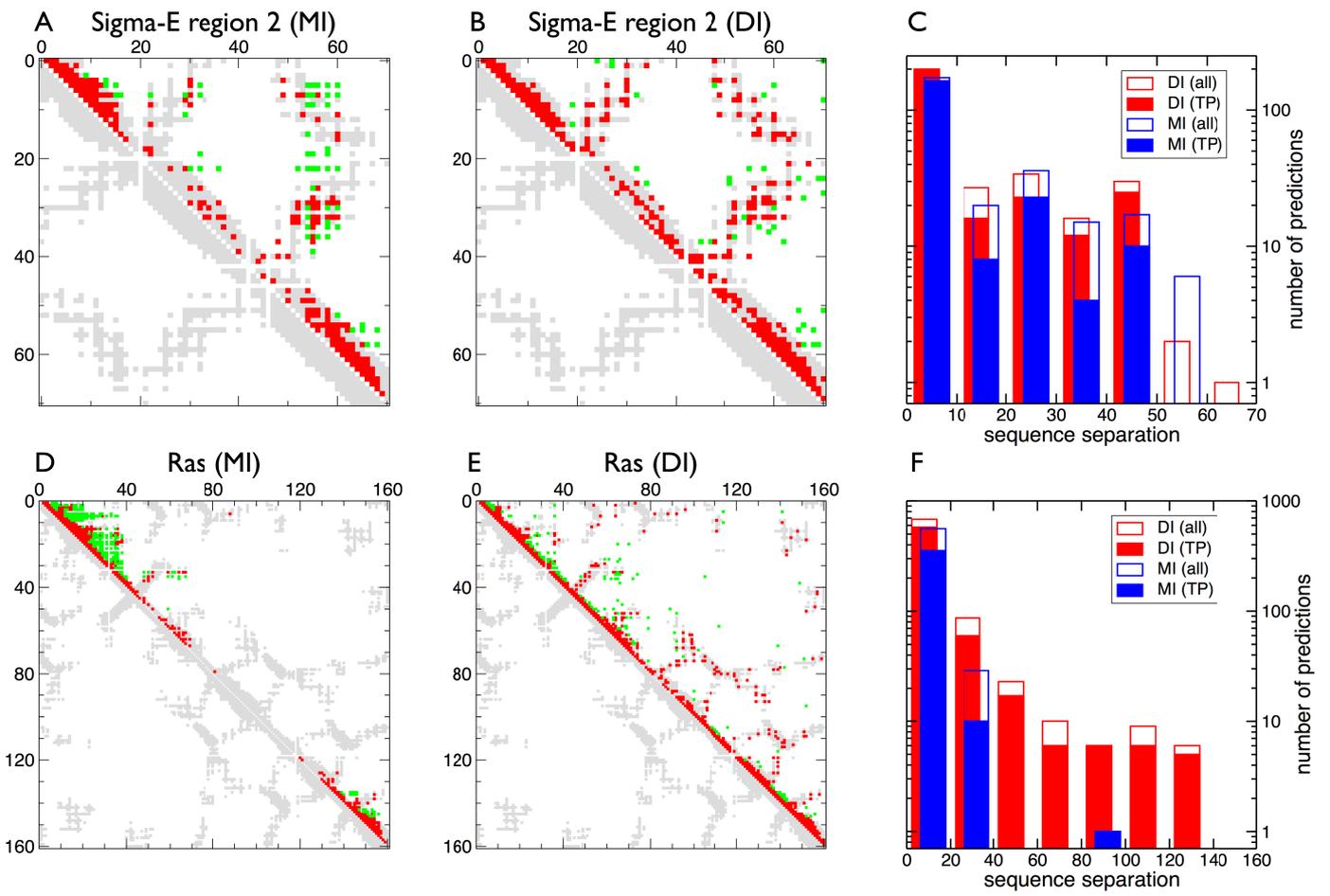